\documentclass[galaxies,article,submit,moreauthors,10pt,a4paper]{mdpi}
\usepackage{amsmath,amssymb,graphicx,paralist}

\firstpage{1}
\articlenumber{x}
\doinum{10.3390/------}
\pubvolume{xx}
\pubyear{2016}
\copyrightyear{2016}
\externaleditor{Academic Editor: name}
\history{Received: date; Accepted: date; Published: date}

\Title{Self-gravitating Bose-Einstein condensates and the Thomas-Fermi approximation}
\Author{Viktor T. Toth$^\dagger$}
\address{$^\dagger$\quad Ottawa ON, Canada, vttoth@vttoth.com}

\abstract{Self-gravitating Bose-Einstein condensates (BEC) have been proposed in various astrophysical contexts, including Bose-stars and BEC dark matter halos. These systems are described by a combination of the Gross-Pitaevskii and Poisson equations (the GPP system). In the analysis of these hypothetical objects, the Thomas-Fermi (TF) approximation is widely used. This approximation is based on the assumption that in the presence of a large number of particles, the kinetic term in the Gross-Pitaevskii energy functional can be neglected, yet it is well known that this assumption is violated near the condensate surface. We also show that the total energy of the self-gravitating condensate in the TF-approximation is positive.  The stability of a self-gravitating system is dependent on the total energy being negative. Therefore, the TF-approximation is ill suited to formulate initial conditions in numerical simulations. As an alternative, we offer an approximate solution of the full GPP system.}

\keyword{Bose-Einstein condensates; Dark matter halos}

\begin{document}

\section{Introduction}

Self-gravitating Bose-Einstein condensates (BECs) have been proposed in various astrophysical contexts, including Bose-stars \cite{1968PhRv..172.1331K,1969PhRv..187.1767R,1986PhRvL..57.2485C,Cha2012} and BEC dark matter halos \cite{1996PhRvD..53.2236L,2007JCAP...06..025B,2011PhRvD..84d3531C,2011PhRvD..84d3532C,2000NewA....5..103G,2003PhRvD..68b3511A,Matos2013}.

Nonrelativistic, dilute BECs are well-described by the Gross-Pitaevskii equation (GPE). In its time-independent form, when kinetic energy can be neglected, the GPE has an approximate solution in the form of the Thomas-Fermi (TF) approximation.

Self-gravitating systems can also be described using the GPE. The long-range gravitational interaction is represented by the external potential. Such systems, referred to in the literature as GPE-Poisson or GPN (Gross-Pitaevskii-Newton) systems, are extensively studied (a few recent examples of interest include \cite{Klaus2015,2014PhRvD..90j3526K,2014NJPh...16k5007B,2014NJPh...16g5005G}).

When applied to self-gravitating systems, the TF-approximation was found to describe an unstable system \cite{Guzman2013}. This also agrees with this author's experience with numerical simulation code, which failed to yield stable self-gravitating systems when the TF-approximation was used as an initial estimate of the condensate density.

In the present paper, it is shown that this instability is the consequence of the TF-approximation, specifically the well-known issue of the divergence of its kinetic energy \cite{Pitaevskii1996,Pethick2008,2011PhRvD..84d3531C,2011PhRvD..84d3532C}. To overcome this issue, we examine the time-independent GPE and attempt to solve it numerically without truncations. We develop a new approximation that has the desirable property that the system has negative total energy and it is stable. This approximation has since been incorporated into our simulation code for self-gravitating GPE-Poisson systems.

\section{Discussion}

A self-gravitating Bose-Einstein condensate is described by a combination of the Gross-Pitaevskii equation and Poisson's equation for gravity. In units such that the BEC particle mass is $m=1$ and also $\hbar=1$, the time-independent Gross-Pitaevskii equation can be written in an attractively simple form:
\begin{align}
-\frac{1}{2}\nabla^2\Psi+(V+c|\Psi|^2-\mu)\Psi&=0,\label{eq:GPE}
\end{align}
where $\Psi$ is the BEC wavefunction, $V$ is the gravitational potential, $c$ is the BEC coupling coefficient and $\mu$ is the chemical potential, the presence of which guarantees the conservation of energy. We normalize the wavefunction such that the number of particles is $\int_V|\Psi|^2=N$.

The energy functional from which with the GPE~(\ref{eq:GPE}) can be derived using the variational principle (cf. \cite{Wang2001,Pethick2008}; note the additional factor of $1/2$ in front of $V$, required to avoid double counting the gravitational potential energy between two regions of the condensate as $V$ is itself a function of $|\Psi|^2$) is given by 
\begin{align}
{\cal E}&=\frac{1}{2}|\nabla\Psi|^2+\left(\frac{1}{2}V-\mu\right)|\Psi|^2+\frac{1}{2}c|\Psi|^4.\label{eq:E}
\end{align}

In the Thomas-Fermi (TF) approximation \cite{2007JCAP...06..025B,2011PhRvD..84d3531C,2011PhRvD..84d3532C}, kinetic energy is neglected. Therefore, the time-independent GPE takes on the following simplified form:
\begin{align}
(V+c|\Psi|^2-\mu)\Psi\simeq 0.
\end{align}
If $V$ is not dependent on $\Psi$, this equation can be solved directly for $|\Psi|^2$:
\begin{align}
|\Psi|^2\simeq\frac{\mu-V}{c}.
\end{align}
Moreover, if we require the wavefunction to vanish at infinity, we must have
\begin{align}
\mu-V\rightarrow 0
\end{align}
at infinity.

If $V$ is dependent on $\Psi$, the situation becomes somewhat more complicated. In particular, in the GPP system, the relationship between $V$ and $\Psi$ is given by Poisson's equation:
\begin{align}
\nabla^2V=4\pi G|\Psi|^2,\label{eq:Poisson}
\end{align}
where $G$ is the gravitational constant. Solving the GPE~(\ref{eq:GPE}) for $V$ in the TF-approximation,
\begin{align}
V\simeq\mu-c|\Psi|^2,
\end{align}
and substituting this solution back into Poisson's equation (\ref{eq:Poisson}), we get
\begin{align}
\nabla^2(\mu-c|\Psi|^2)\simeq 4\pi G|\Psi|^2.
\end{align}
If $\mu={\rm const.}$, we are left with
\begin{align}
\left[\nabla^2+\frac{4\pi G}{c}\right]|\Psi|^2\simeq 0,
\end{align}
which is an homogeneous Helmholtz-type equation for $|\Psi|^2$, spherically symmetric solutions of which are
\begin{align}
|\Psi|^2\simeq C_1\frac{\sin kr}{r}+C_2\frac{\cos kr}{r},
\end{align}
where $k^2=4\pi G/c$, while $C_1$ and $C_2$ are integration constants. To avoid solutions that are singular at the origin $r=0$, we must set $C_2=0$. On the other hand, $\sin kr/r$ (and thus, $|\Psi|^2$) vanishes at $r=\pi/k$. Therefore, we set $r_0=\pi/k$ as the radius of the condensate. This determines $C_1$ since we require that
\begin{align}
\int_V C_1\frac{\sin kr}{r}~dV=N.
\end{align}
This integral can be readily evaluated:
\begin{align}
\int_V C_1\frac{\sin kr}{r}~dV=4\pi C_1\int_0^{r_0} r\sin\frac{\pi r}{r_0}~dr=4C_1r_0^2,
\end{align}
hence $C_1=N/4r_0^2$. Therefore, the TF-approximation for the GPP is given by the Lane-Emden type solution
\begin{align}
|\Psi|^2=\frac{N}{4r_0^2}\frac{\sin\left(\pi r/r_0\right)}{r},\label{eq:LE}
\end{align}
for $0\leq r \leq r_0=\sqrt{\pi c/4 G}$.

Given $|\Psi|^2$, we can solve the GPE~(\ref{eq:GPE}) for $V$:
\begin{align}
V=\mu-c|\Psi|^2=\mu-\frac{cN}{4r_0^2}\frac{\sin\left(\pi r/r_0\right)}{r},
\end{align}
again for $0\leq r \leq r_0$.

At $r_0$ and beyond, the condensate vanishes, and the gravitational potential becomes that of a point mass $M$ (where $M=Nm$ is the total mass of the condensate), i.e., $V=-GN/r$ ($r_0\leq r$). At the boundary, these two forms must agree. This can be achieved by setting
\begin{align}
\mu=-\frac{GN}{r_0}.
\end{align}
This clarifies the role of the chemical potential in the case of the GPP system in the TF-approximation: its presence ensures that the gravitational potential takes on the standard form outside the condensate and vanishes at infinity.

The energy density of the time-independent GPE in the Thomas-Fermi limit is given by
\begin{align}
{\cal E}\simeq\left(\frac{1}{2}V-\mu\right)|\Psi|^2+\frac{1}{2}c|\Psi|^4,
\end{align}
or, after substituting the solution for $V$ from the GPE~(\ref{eq:GPE}),
\begin{align}
{\cal E}\simeq-c|\Psi|^4+\frac{1}{2}c|\Psi|^4=-\frac{1}{2}c|\Psi|^4.
\end{align}
\begin{align}
{\cal E}\simeq-\frac{1}{2}\mu|\Psi|^2=\frac{GN}{r_0}|\Psi|^2.
\end{align}
To find the total energy, we integrate over the condensate volume:
\begin{align}
E&=\int_V{\cal E}~dV=4\pi\int_0^{r_0}r^2{\cal E}~dr\nonumber\\
&\simeq \frac{\pi GN^2}{r_0^3}\int_0^{r_0}r\sin(\pi r/r_0)~dr=\frac{GN^2}{r_0},
\end{align}
or, after restoring units,
\begin{align}
E\simeq \frac{Gm^2N^2}{r_0}.
\end{align}
The positive sign of the total energy implies that the solution for a self-gravitating BEC using the TF-approximation is inherently unstable.

This result is based on the assumption that the kinetic energy can be neglected. Now that we have an explicit solution for $|\Psi|^2$, this assumption can be verified by direct substitution into the energy functional~(\ref{eq:E}). When we do so we find that, using the solution given by Eq.~(\ref{eq:LE}), the condensate kinetic energy,
\begin{align}
{\rm KE}&=\int\frac{1}{2}|\nabla\Psi|^2~dV=4\pi\int_0^{r_0}\frac{r^2}{2}|\nabla\Psi|^2~dr\nonumber\\
&
=\frac{\pi N}{2r_0^2}\int_0^{r_0}r^2\left|\nabla\sqrt{\frac{\sin(\pi r/r_0)}{r}}\right|^2~dr\label{eq:KE}
\end{align}
is divergent for any $r_0>0$.

The implication of this divergence is that the assumption behind the TF-approximation, namely that the kinetic term in the GPE~(\ref{eq:GPE}) can be neglected, is maximally violated.

The divergence of the kinetic energy density (represented by the integrand in Eq.~({\ref{eq:KE})) in the Thomas-Fermi approximation, near the condensate surface, and the logarithmic divergence of the total kinetic energy are well known\footnote{The author wishes to thank the anonymous referees and the Academic Editor of {\em Galaxies} for stressing this point.} \cite{Pitaevskii1996,Pethick2008,2011PhRvD..84d3531C,2011PhRvD..84d3532C}. However, in the case of self-gravitating systems, as we have seen, the problem gets worse: the total energy of a self-gravitating condensate is positive, hence the condensate is gravitationally unstable. This has especially important implications for numerical simulations of self-gravitating Bose-Einstein condensates that use this approximation to model the initial state (see, e.g., \cite{Madarassy2013,Guzman2013}). Not only is the magnitude of the total kinetic energy ill-defined (which makes it dependent on nonphysical simulation parameters, such as the numerical integration step size or even small rounding errors), the positive total energy is especially troublesome, as the stability of a self-gravitating system is dependent on $E<0$.

This finding agrees with the author's experience using numerical simulation code \cite{Madarassy2013} that was designed to model the (time-dependent) GPP system, using the TF-approximation to model the initial condensate density. It was perplexing that versions of the code ran differently in different programming environments (e.g., single vs. double precision, FORTRAN vs. C), processors (Intel x86 vs. GPGPU) and operating systems (Linux vs. Windows). This is clearly not permissible: the results, apart from accuracy and rounding issues, should not be dependent on such factors. Indeed, the present study arose as a result of systematically analyzing the failure of these algorithms to produce consistent results.

Numerically stable simulations require an initial state that is not dependent on nonphysical parameters and has a well-defined total energy that is consistent with stability conditions. This is clearly not the case when the TF-approximation is used for a self-gravitating condensate. Therefore, we now aim to find an approximate solution of the GPP system in the spherically symmetric case without resorting to the TF-approximation (see also \cite{2011PhRvD..84d3531C} for another approximation and \cite{2011PhRvD..84d3532C} for a numeric solution in terms of hydrodynamics using the Madelung-representation). Assuming spherical coordinates and a spherically symmetric condensate, the GPE~(\ref{eq:GPE}) becomes an equation of a single independent variable $r$. Let us denote $\partial/\partial r = \partial_r$ for brevity, while noting the form of the Laplacian in spherical coordinates, $\nabla^2=\partial^2_r+(2/r)\partial_r$. We can then write the GPE~(\ref{eq:GPE}) as
\begin{align}
\partial^2_r\Psi+\frac{2}{r}\partial_r\Psi-2V\Psi-2c|\Psi|^2\Psi+2\mu\Psi=0,
\end{align}
whereas Poisson's equation (\ref{eq:Poisson}) becomes:
\begin{align}
\partial^2_r V+\frac{2}{r}\partial_r V=4\pi G|\Psi|^2.
\end{align}
Let us now consider writing the wavefunction as
\begin{align}
\Psi(r)=|\Psi|(r)e^{i\phi(r)},
\end{align}
in which case
\begin{align}
\hskip -0.15in d\Psi&=e^{i\phi}\left[d|\Psi| + i|\Psi|d\phi\right],\\
\hskip -0.15in d^2\Psi&=e^{i\phi}\{d^2|\Psi| - |\Psi|(d\phi)^2 + i[2d|\Psi|d\phi + |\Psi|d^2\phi]\},
\end{align}
and the GPE~(\ref{eq:GPE}) can be rewritten, after dividing through by $e^{i\phi}$, as
\begin{align}
\partial_r^2\Psi|-|\Psi|(\partial_r\phi)^2+\frac{2}{r}\partial_r|\Psi|
+i\left(2\partial_r|\Psi|\partial_r\phi+|\Psi|\partial_r^2\phi+\frac{2}{r}|\Psi|\partial_r\phi\right)
-2V|\Psi|-2c|\Psi|^3+2\mu|\Psi|&=0.
\end{align}
Since $|\Psi|$, $\phi$, $V$, $c$ and $\mu$ are all real, the real and imaginary parts of this equation can be separated:
\begin{align}
\hskip -2.5em\partial_r^2|\Psi|+\frac{2}{r}\partial_r|\Psi|-\left[\left(\partial_r\phi\right)^2+2V+2c|\Psi|^2-2\mu\right]|\Psi|&=0,\\
\partial_r^2\phi+2\left[\partial_r\ln|\Psi|+\frac{1}{r}\right]\partial_r\phi&=0.\label{eq:phi}
\end{align}
Equation~(\ref{eq:phi}) can be integrated:
\begin{align}
\partial_r\phi=\frac{C}{r^2|\Psi|^2},
\end{align}
where $C$ is an integration constant with the dimensions of $r|\Psi|^2$. It seems that $C=0$ is not only a valid choice but the only choice that does not result in $\phi$ becoming singular at the origin (assuming the wavefunction does not vanish at the origin). Therefore, $\phi={\rm const.}$

\begin{figure}[t]
\begin{center}{\includegraphics[width=0.5\linewidth]{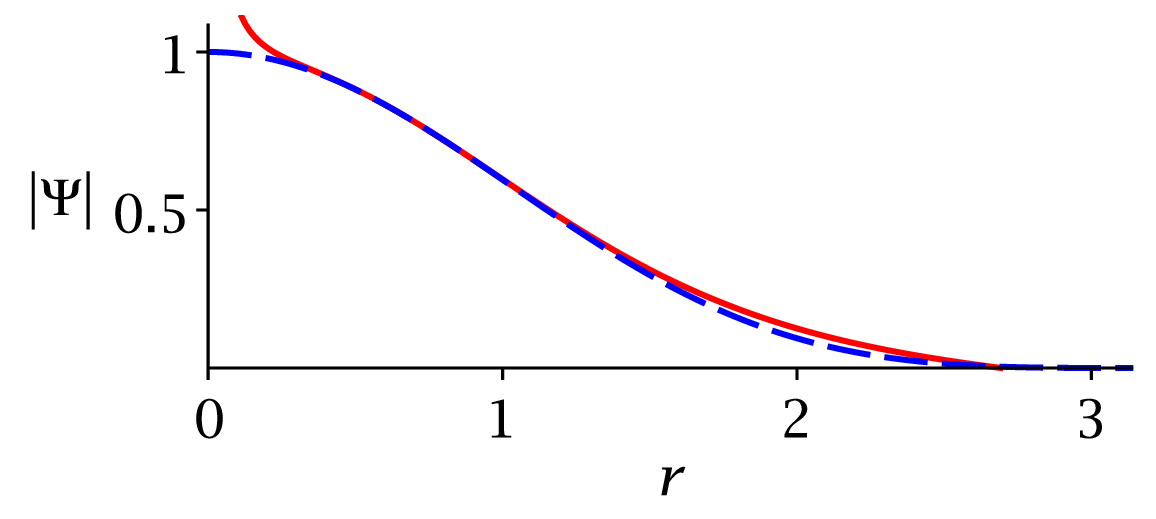}}\end{center}\vskip -0.25in
\caption{Numerical solution (solid red line) of Eq.~(\ref{eq:psi}), giving the norm of the wavefunction $|\Psi|$ of a spherically symmetric self-gravitating BEC  as a function of radius $r$, compared to the approximation (dashed blue line) provided in Eq.~(\ref{eq:ini}).\label{fig:linsol}}
\end{figure}

Under these circumstances, the GPE~(\ref{eq:GPE}) will read
\begin{align}
\partial_r^2|\Psi|+\frac{2}{r}\partial_r|\Psi|-\left[2V+2c|\Psi|^2-2\mu\right]|\Psi|&=0,
\end{align}
which is readily solvable for $V$ algebraically:
\begin{align}
\hskip -0.1in
V=\frac{\partial_r^2|\Psi|}{2|\Psi|}+\frac{\partial_r|\Psi|}{r|\Psi|}-c|\Psi|^2+\mu.
\end{align}
This result can be substituted back into Poisson's equation, yielding a fourth-order ordinary differential equation in $|\Psi|$:
\begin{align}
\partial_r^4|\Psi|-\frac{2\partial|\Psi|\partial_r^3|\Psi|}{|\Psi|}+\frac{4\partial_r^3|\Psi|}{r}-\frac{(\partial_r^2|\Psi|)^2}{|\Psi|}-4c|\Psi|^2\partial_r^2|\Psi|+\frac{2(\partial_r|\Psi|)^2\partial_r^2|\Psi|}{|\Psi|^2}-{}&\frac{8\partial_r|\Psi|\partial_r^2|\Psi|}{r|\Psi|}\nonumber\\
+\frac{4(\partial_r|\Psi|)^3}{r|\Psi|^2}-4c|\Psi|(\partial_r|\Psi|)^2-\frac{8c|\Psi|^2\partial_r|\Psi|}{r}-{}&8\pi G|\Psi|^3=0.\label{eq:psi}
\end{align}

\begin{figure}[t]
\begin{center}{\includegraphics[trim=120 50 100 0,width=0.4\linewidth]{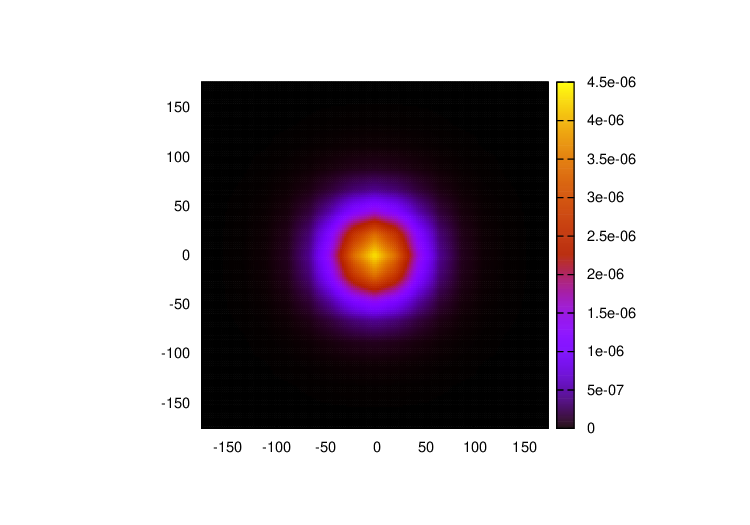}}\end{center}\vskip 1em\caption{Density cross-section of a stable simulated $1~M_\odot$, $r\simeq 50$~km Bose-star (or stellar core) after approximately 300,000 numerical iterations that corresponds to 3 seconds \cite{Madarassy2014}. (For comparison, the period of a circular orbit at $r=50$~km is approximately 0.006~s.) Axes are in km, density is in units of $10^{30}~{\rm kg}/{\rm km}^3\simeq 0.5~M_{\odot}/{\rm km}^3$. NB: The spatial grid used in this simulation is low resolution ($60\times 60\times 60$) and image smoothing was used to improve the presentation quality.\label{fig:star}}
\end{figure}

This equation can be solved numerically. Given that it is a fourth-order homogeneous differential equation in $|\Psi|$, it has a very large solution space, parameterized by boundary or initial conditions, such as the values of $|\Psi|$ and its first three derivatives at some value of $r$. A hint for a suitable solution comes from numerical simulation \cite{Madarassy2013}, where we find that apparently stable nonrotating spherically symmetric solutions converge on $|\Psi|^2\propto[\sin(r/r_0)/(r/r0)]^{2\alpha}$, with $3\lesssim\alpha\lesssim4$. This approximate solution has many desirable properties. It is smooth in the interval $0\ge r/r_0\ge\pi$. The corresponding kinetic energy (\ref{eq:KE}) is finite in the same interval. Moreover, it is possible to compute the condensate mass, which is given by
\begin{align}
M(r)=\int_V|\Psi|^2~dV=4\pi\int_0^rr'^2|\Psi(r')|^2~dr',
\end{align}
and this, too, is finite and well-behaved. Therefore, we find that the following initial approximation for the magnitude of the BEC-Poisson wavefunction:
\begin{equation}
|\Psi|\propto\left|\frac{\sin r/r_0}{r/r_0}\right|^3,
\label{eq:ini}
\end{equation}
agrees well with a numerical solution (except for very small values of $r$), as shown in Fig.~\ref{fig:linsol}. Furthermore, this choice yields a corresponding solution of Eq.~(\ref{eq:Poisson}) for the gravitational potential that is finite, negative, and vanishes at infinity, as expected. The stability of these solutions is confirmed by numerical simulation of a condensate using Eq.~(\ref{eq:ini}) as an initial approximation. An example result is shown in Fig.~\ref{fig:star}.

\section{Conclusions}

We have demonstrated that when the Thomas-Fermi approximation is used to describe self-gravitating Bose-Einstein condensates in astrophysical contexts, the resulting systems have divergent (positive) total energy and are unstable. However, this behavior is a specific consequence that arises from the use of the TF-approximation; GPE-Poisson systems are not inherently unstable. By investigating the untruncated Gross-Pitaevskii equation, we found that the total energy is, in fact negative. Furthermore, using an approximate numerical solution as a guide, we developed a simple approximation formula that can be used to provide an initial density estimate for BECs in numerical simulations. These simulations can be applied to describe, e.g., BEC dark matter stars or stellar cores; these results and analysis will be reported elsewhere \cite{Madarassy2014} as they become available.

\acknowledgments{I thank Tiberiu Harko and E. J. M. Madarassy for especially valuable discussions.}

\conflictofinterests{The author declares no conflict of interest.}

\bibliographystyle{mdpi}

\bibliography{refs}

\begin{thebibliography}{-------}
\providecommand{\natexlab}[1]{#1}

\bibitem[{Kaup}(1968)]{1968PhRv..172.1331K}
{Kaup}, D.J.
\newblock {Klein-Gordon Geon}.
\newblock {\em Physical Review} {\bf 1968}, {\em 172},~1331--1342.

\bibitem[{Ruffini} and {Bonazzola}(1969)]{1969PhRv..187.1767R}
{Ruffini}, R.; {Bonazzola}, S.
\newblock {Systems of Self-Gravitating Particles in General Relativity and the
  Concept of an Equation of State}.
\newblock {\em Physical Review} {\bf 1969}, {\em 187},~1767--1783.

\bibitem[{Colpi} \em{et~al.}(1986){Colpi}, {Shapiro}, and
  {Wasserman}]{1986PhRvL..57.2485C}
{Colpi}, M.; {Shapiro}, S.L.; {Wasserman}, I.
\newblock {Boson stars - Gravitational equilibria of self-interacting scalar
  fields}.
\newblock {\em Physical Review Letters} {\bf 1986}, {\em 57},~2485--2488.

\bibitem[{Chavanis} and {Harko}(2012)]{Cha2012}
{Chavanis}, P.H.; {Harko}, T.
\newblock {Bose-Einstein condensate general relativistic stars}.
\newblock {\em \prd} {\bf 2012}, {\em 86},~064011,
  \href{http://xxx.lanl.gov/abs/arXiv:1108.3986 [astro-ph.SR]}{{\normalfont
  [arXiv:astro-ph.SR/arXiv:1108.3986 [astro-ph.SR]]}}.

\bibitem[{Lee} and {Koh}(1996)]{1996PhRvD..53.2236L}
{Lee}, J.W.; {Koh}, I.G.
\newblock {Galactic halos as boson stars}.
\newblock {\em \prd} {\bf 1996}, {\em 53},~2236--2239,
  \href{http://xxx.lanl.gov/abs/{arXiv:hep-ph/9507385}}{{\normalfont
  [{arXiv:hep-ph/9507385}]}}.

\bibitem[{B{\"o}hmer} and {Harko}(2007)]{2007JCAP...06..025B}
{B{\"o}hmer}, C.G.; {Harko}, T.
\newblock {Can dark matter be a Bose Einstein condensate?}
\newblock {\em \jcap} {\bf 2007}, {\em 6},~025,
  \href{http://xxx.lanl.gov/abs/{arXiv:0705.4158 [astro-ph]}}{{\normalfont
  [{arXiv:0705.4158 [astro-ph]}]}}.

\bibitem[{Chavanis}(2011)]{2011PhRvD..84d3531C}
{Chavanis}, P.H.
\newblock {Mass-radius relation of Newtonian self-gravitating Bose-Einstein
  condensates with short-range interactions. I. Analytical results}.
\newblock {\em \prd} {\bf 2011}, {\em 84},~043531,
  \href{http://xxx.lanl.gov/abs/1103.2050}{{\normalfont
  [arXiv:astro-ph.CO/1103.2050]}}.

\bibitem[{Chavanis} and {Delfini}(2011)]{2011PhRvD..84d3532C}
{Chavanis}, P.H.; {Delfini}, L.
\newblock {Mass-radius relation of Newtonian self-gravitating Bose-Einstein
  condensates with short-range interactions. II. Numerical results}.
\newblock {\em \prd} {\bf 2011}, {\em 84},~043532,
  \href{http://xxx.lanl.gov/abs/arXiv:1103.2054 [astro-ph.CO]}{{\normalfont
  [arXiv:astro-ph.CO/arXiv:1103.2054 [astro-ph.CO]]}}.

\bibitem[{Goodman}(2000)]{2000NewA....5..103G}
{Goodman}, J.
\newblock {Repulsive dark matter}.
\newblock {\em \na} {\bf 2000}, {\em 5},~103--107,
  \href{http://xxx.lanl.gov/abs/{arXiv:astro-ph/0003018}}{{\normalfont
  [{arXiv:astro-ph/0003018}]}}.

\bibitem[{Arbey} \em{et~al.}(2003){Arbey}, {Lesgourgues}, and
  {Salati}]{2003PhRvD..68b3511A}
{Arbey}, A.; {Lesgourgues}, J.; {Salati}, P.
\newblock {Galactic halos of fluid dark matter}.
\newblock {\em \prd} {\bf 2003}, {\em 68},~023511,
  \href{http://xxx.lanl.gov/abs/{arXiv:astro-ph/0301533}}{{\normalfont
  [{arXiv:astro-ph/0301533}]}}.

\bibitem[Su{\'a}rez \em{et~al.}(2013)Su{\'a}rez, Robles, and Matos]{Matos2013}
Su{\'a}rez, A.; Robles, V.; Matos, T.
\newblock {A Review on the Scalar Field/ Bose-Einstein Condensate Dark Matter
  Model}.
\newblock  {Proceedings of the IV International Meeting on Gravitation and
  Cosmology},  2013, Vol.~38, {\em Astrophysics and Space Science Proceedings},
   \href{http://xxx.lanl.gov/abs/{arXiv:1302.0903 [astro-ph.CO]}}{{\normalfont
  [{arXiv:1302.0903 [astro-ph.CO]}]}}.
\newblock {Chapter 9}.

\bibitem[{Schroven} \em{et~al.}(2015){Schroven}, {List}, and
  {L{\"a}mmerzahl}]{Klaus2015}
{Schroven}, K.; {List}, M.; {L{\"a}mmerzahl}, C.
\newblock {Stability of self-gravitating Bose-Einstein condensates}.
\newblock {\em \prd} {\bf 2015}, {\em 92},~124008,
  \href{http://xxx.lanl.gov/abs/1507.06122}{{\normalfont
  [arXiv:gr-qc/1507.06122]}}.

\bibitem[{K{\"u}hnel} and {Rampf}(2014)]{2014PhRvD..90j3526K}
{K{\"u}hnel}, F.; {Rampf}, C.
\newblock {Astrophysical Bose-Einstein condensates and superradiance}.
\newblock {\em \prd} {\bf 2014}, {\em 90},~103526,
  \href{http://xxx.lanl.gov/abs/1408.0790}{{\normalfont
  [arXiv:gr-qc/1408.0790]}}.

\bibitem[{Bahrami} \em{et~al.}(2014){Bahrami}, {Gro{\ss}ardt}, {Donadi}, and
  {Bassi}]{2014NJPh...16k5007B}
{Bahrami}, M.; {Gro{\ss}ardt}, A.; {Donadi}, S.; {Bassi}, A.
\newblock {The Schr{\"o}dinger-Newton equation and its foundations}.
\newblock {\em New Journal of Physics} {\bf 2014}, {\em 16},~115007,
  \href{http://xxx.lanl.gov/abs/1407.4370}{{\normalfont
  [arXiv:quant-ph/1407.4370]}}.

\bibitem[{Giulini} and {Gro{\ss}ardt}(2014)]{2014NJPh...16g5005G}
{Giulini}, D.; {Gro{\ss}ardt}, A.
\newblock {Centre-of-mass motion in multi-particle Schr{\"o}dinger-Newton
  dynamics}.
\newblock {\em New Journal of Physics} {\bf 2014}, {\em 16},~075005,
  \href{http://xxx.lanl.gov/abs/1404.0624}{{\normalfont
  [arXiv:gr-qc/1404.0624]}}.

\bibitem[{Guzm{\'a}n} \em{et~al.}(2013){Guzm{\'a}n}, {Lora-Clavijo},
  {Gonz{\'a}lez-Avil{\'e}s}, and {Rivera-Paleo}]{Guzman2013}
{Guzm{\'a}n}, F.S.; {Lora-Clavijo}, F.D.; {Gonz{\'a}lez-Avil{\'e}s}, J.J.;
  {Rivera-Paleo}, F.J.
\newblock {Stability of BEC galactic dark matter halos}.
\newblock {\em \jcap} {\bf 2013}, {\em 9},~34,
  \href{http://xxx.lanl.gov/abs/1308.4925}{{\normalfont
  [arXiv:astro-ph.CO/1308.4925]}}.

\bibitem[Dalfovo \em{et~al.}(1996)Dalfovo, Pitaevskii, and
  Stringari]{Pitaevskii1996}
Dalfovo, F.; Pitaevskii, L.P.; Stringari, S.
\newblock {The Condensate Wave Function of a Trapped Atomic Gas}.
\newblock {\em J. Res. Natl. Inst. Stand. Tech.} {\bf 1996}, {\em
  101},~537--544.

\bibitem[{Pethick, C.~J. and Smith, H.}(2008)]{Pethick2008}
{Pethick, C.~J. and Smith, H.}.
\newblock {\em {Bose-Einstein condensation in dilute gases}}, second ed.;
  Cambridge University Press,  2008.

\bibitem[{Wang}(2001)]{Wang2001}
{Wang}, X.Z.
\newblock {Cold Bose stars: Self-gravitating Bose-Einstein condensates}.
\newblock {\em \prd} {\bf 2001}, {\em 64},~124009.

\bibitem[{Madarassy} and {Toth}(2013)]{Madarassy2013}
{Madarassy}, E.J.M.; {Toth}, V.T.
\newblock {Numerical simulation code for self-gravitating Bose{\ndash}Einstein
  condensates}.
\newblock {\em Computer Physics Communications} {\bf 2013}, {\em
  184},~1339--1343,  \href{http://xxx.lanl.gov/abs/1207.5249}{{\normalfont
  [arXiv:astro-ph.GA/1207.5249]}}.

\bibitem[{Madarassy} and {Toth}(2015)]{Madarassy2014}
{Madarassy}, E.J.M.; {Toth}, V.T.
\newblock {Evolution and dynamical properties of Bose-Einstein condensate dark
  matter stars}.
\newblock {\em \prd} {\bf 2015}, {\em 91},~044041,
  \href{http://xxx.lanl.gov/abs/1412.7152}{{\normalfont
  [arXiv:hep-ph/1412.7152]}}.

\end{thebibliography}

\end{document}